\documentclass[12pt]{elsart}

\usepackage{amsmath}
\usepackage{amsfonts}
\usepackage{graphicx}

\begin{document}

\begin{frontmatter}

\title{Some features of the statistical complexity, Fisher-Shannon information, 
and Bohr-like orbits in the Quantum Isotropic Harmonic Oscillator}

\author[jsr]{Jaime Sa\~{n}udo} and
\ead{jsr@unex.es}
\author[rlr]{Ricardo L\'{o}pez-Ruiz}
\ead{rilopez@unizar.es}

\address[jsr]{
Departamento de F\'isica, Facultad de Ciencias, \\
Universidad de Extremadura, E-06071 Badajoz, Spain}

\address[rlr]{
DIIS and BIFI, Facultad de Ciencias, \\
Universidad de Zaragoza, E-50009 Zaragoza, Spain}


\begin{abstract}
The Fisher-Shannon information and a statistical measure of complexity 
are calculated in the position and momentum spaces for the wave functions of the 
quantum isotropic harmonic oscillator.
We show that these magnitudes are independent of the strength of the 
harmonic potential. Moreover, for each level of energy, it is found that these 
two indicators take their minimum values on the orbitals that correspond to the 
classical (circular) orbits in the Bohr-like quantum image, 
just those with the highest orbital angular momentum. 
\end{abstract}


\end{frontmatter}

\maketitle

In recent years, the study of statistical magnitudes on quantum systems has taken
an increasing interest \cite{gadre1987,panos2005}. The tools that have been developed
in information and complexity theories, for instance, Fisher and Shannon informations
\cite{fisher1925,shannon1948,dembo1991}, and statistical 
indicators of complexity \cite{lopez1995,lopez2002} 
have been calculated for several systems under different approaches 
\cite{gadre1985,dehesa1994,dehesa2004,dehesa2005,panos2007,sen2007,sen2008}.       
The probability densities characterizing the state of a quantum system
are  defined in the position and the momentum spaces \cite{landau1981,galindo1991}.
From here, the calculation of all those statistical indicators can be performed 
with a low computational cost.

The dependence of these quantities on the quantum numbers of the system
can reflect the hierarchical organization of that quantum system.
Even for states with the same energy 
it is possible to have different values of these statistical magnitudes. 
Take, for instance, the $H$-atom.
It has been shown \cite{sanudo2008} that for a given energy the minimum values 
of the Fisher-Shannon information and statistical complexity are reached 
for the highest allowed orbital angular momentum for that energy. 
This means that a variational process on these statistical measures 
can select just those orbitals that in the pre-quantum image are the Bohr-like orbits.

Following this insight, it is our aim in the present work to analyze if the
above described behavior of these statistical measures can be also found in the case
of the isotropic quantum harmonic oscillator. 

Let us start by recalling the three-dimensional non-relativistic wave functions 
of this system when the potential energy is written as $V(r)=\lambda^2r^2/2$, where 
$\lambda$ is a positive real constant expressing the potential strength. 
Atomic units are used through the text.
The wave functions in position space ($\vec{r}=(r,\Omega)$, with $r$ the radial distance and
$\Omega$ the solid angle) are:
\begin{equation}
\Psi_{n,l,m}(\vec{r})= R_{n,l}(r)\; Y_{l,m}(\Omega)\;,
\label{eq:Psi}
\end{equation}
where $R_{n,l}(r)$ is the radial part and $Y_{l,m}(\Omega)$ is the spherical harmonic 
of the quantum state determined by the quantum numbers $(n,l,m)$. The radial part is expressed 
as \cite{dehesa1994}
\begin{equation}
R_{n,l}(r)= \left[{2\;n!\;\lambda^{l+3/2}\over \Gamma(n+l+3/2)}\right]^{1/2}\;
r^l\;e^{-{\lambda \over 2}r^2}\; L_{n}^{l+1/2}(\lambda r^2)\;,
\label{eq:R}
\end{equation}
being $L_{\alpha}^{\beta}(t)$ the associated Laguerre polynomials.
The levels of energy are given by 
\begin{equation}
E_{n,l}=\lambda (2n+l+3/2) = \lambda (e_{n,l}+3/2),
\label{eq:E}
\end{equation}
where $n=0,1,2,\cdots$ and $l=0,1,2,\cdots$.
Let us observe that $e_{n,l}=2n+l$.
Thus, different pairs of $(n,l)$ can give the same $e_{n,l}$,
and then the same energy $E_{n,l}$.

The wave functions in momentum space ($\vec{p}=(p,\hat{\Omega})$, 
with $p$ the momentum modulus and $\hat{\Omega}$ the solid angle) are:
\begin{equation}
\hat{\Psi}_{n,l,m}(\vec{p})= \hat{R}_{n,l}(p)\; Y_{l,m}(\hat{\Omega})\;,
\label{eq:Psih}
\end{equation}
where the radial part $\hat{R}_{n,l}(p)$ is now given by the expression \cite{dehesa1994}
\begin{equation}
\hat{R}_{n,l}(p)= \left[{2\;n!\;\lambda^{-l-3/2}\over \Gamma(n+l+3/2)}\right]^{1/2}\;
p^l\;e^{-{p^2\over 2\lambda}}\; L_{n}^{l+1/2}({p^2/\lambda})\;.
\label{eq:Rh}
\end{equation}

Taking the former expressions, the probability density
in position and momentum spaces,
\begin{equation}
\rho_{n,l,m;\,\lambda}(\vec{r})\;=\;\mid\Psi_{n,l,m}(\vec{r})\mid^2\;, \hspace{1cm}
\gamma_{n,l,m;\,\lambda}(\vec{p})\;=\;\mid\hat{\Psi}_{n,l,m}(\vec{p})\mid^2\;,
\label{eq:rho}
\end{equation}
can be explicitly calculated. From these densities, 
the statistical complexity and the Fisher-Shannon information are computed.
We find that these quantities are independent of $\lambda$, the potential strength. 
This is a non trivial property that is shown in the Appendix A.
For this reason, we drop the $\lambda$ subindex in the densities from now and on.  
Also, for the sake of simplicity, the quantum numbers $(n,l,m)$ are omitted in the notation.

First, the measure of complexity $C$ recently introduced by Lopez-Ruiz, Mancini and Calbet
\cite{lopez1995,lopez2001,lopez2002},
the so-called $LMC$ complexity, is defined as
\begin{equation}
C = H\cdot D\;,
\end{equation}
where $H$ represents the information content of the system and $D$ is the
distance from the actual state of the system to some prestablished reference state. 

For our purpose, we take a version used in Ref. \cite{lopez2002}
as quantifier of $H$. This is the simple exponential Shannon entropy,
that in the position and momentum spaces takes the form, respectively,
\begin{equation}
H_r = e^{S_r}\;, \hspace{1cm}
H_p = e^{S_p}\;.
\label{eq:H}
\end{equation}
$S_r$ and $S_p$ are the Shannon information entropies \cite{shannon1948},
\begin{equation}
S_r = -\int\rho(\vec{r})\;\log \rho(\vec{r})\; d\vec{r}\;, \hspace{1cm}
S_p = -\int\gamma(\vec{p})\;\log \gamma(\vec{p})\; d\vec{p}\;.
\end{equation}
We keep for the disequilibrium the form originally introduced in 
Refs. \cite{lopez1995,lopez2002}, that is,
\begin{equation}
D_r = \int\rho^2(\vec{r})\; d\vec{r}\;, \hspace{1cm}
D_p = \int\gamma^2(\vec{p})\; d\vec{p}\;,
\label{eq:D}
\end{equation}
In this manner, the final expressions for $C$ in position and 
momentum spaces are:  
\begin{equation}
C_r = H_r\cdot D_r\;, \hspace{1cm}
C_p = H_p\cdot D_p\;.
\end{equation}
The form of the wave functions, due to the harmonic interaction, 
allows us to show in the Appendix A that these magnitudes, 
$C_r$ and $C_p$, are the same.    

In Fig. 1, $C_r$ (or $C_p$) is plotted as function of the modulus of the
third component $m$, $-l\leq m \leq l$, of the orbital angular momentum $l$ for different $l$ values
with a fixed energy. That is, according to expression (\ref{eq:E}), 
the quantity $e_{n,l}=2n+l$ is constant in each figure.
Fig. 1(a) shows $C_r$ for $e_{n,l}=15$ and Fig. 1(b) shows $C_ r$ for $e_{n,l}=30$.
In both figures, it can be observed that $C_r$ splits in different sets of discrete points.
Each one of these sets is associated to a different $l$ value. 
It is worth to note that the set with the minimum values of $C_r$ corresponds just
to the highest $l$, that is, $l=15$ in Fig. 1(a) and $l=30$ in Fig. 1(b).

Second, other types of statistical measures that maintain the product form of $C$ can be defined.
Let us take, for instance, the Fisher-Shannon information, $P$, that has been also applied
in Refs. \cite{dehesa2004,sen2008,sanudo2008} for quantum systems. This quantity, in the position 
and momentum spaces, is given respectively by   
\begin{equation}
P_r = J_r\cdot I_r\;, \hspace{1cm}
P_p = J_p\cdot I_p\;,
\label{eq:P}
\end{equation}
where the first factor
\begin{equation}
J_r = {1\over 2\pi e}\;e^{2S_r/3}\;, \hspace{1cm}
J_p = {1\over 2\pi e}\;e^{2S_p/3}\;,
\label{eq:J}
\end{equation}
is a version of the exponential Shannon entropy \cite{dembo1991}, 
and the second factor
\begin{equation}
I_r = \int{[\vec{\nabla}\rho(\vec{r})]^2\over \rho(\vec{r})}\; d\vec{r}\;, \hspace{1cm}
I_p = \int{[\vec{\nabla}\gamma(\vec{p})]^2\over \gamma(\vec{p})}\; d\vec{p}\;,
\label{eq:I}
\end{equation}
is the so-called Fisher information measure \cite{fisher1925}, that quantifies the narrowness 
of the probability density. 
Similarly to the behavior of $C_r$ and $C_p$,
we also show in the Appendix A that $P_r=P_p$.

$I_r$ can be analytically obtained in both spaces (position and momentum). 
The results are \cite{dehesa2005}:
\begin{eqnarray}
\hspace{2cm} I_r & = & 4\,(2n+l+3/2-|m|)\,\lambda \, , \label{eq:Ir}\\
\hspace{2cm} I_p & = & 4\,(2n+l+3/2-|m|)\,\lambda^{-1}. \label{eq:Ip}
\end{eqnarray}
Let us note that $I_r$ and $I_p$ depend on $\lambda$,
although the final result for $P_r$ and $P_p$ are non $\lambda$-dependent (see Appendix A).

Fig. 2 shows the calculation of $P$ as function of the modulus of the
third component $m$ for different pairs of $(e_{n,l}=2n+l,l)$ values. 
In Fig. 2(a), $P_r$ (or $P_p$) is plotted for $e_{n,l}=15$, 
and $P_ r$ is plotted for $e_{n,l}=30$ in Fig. 2(b).
Here, $P_r$ also splits in different sets 
of discrete points, showing a behavior parallel to the above signaled for $C$ (Fig. 1). 
Each one of these sets is related with a different $l$ value, 
and the set with the minimum values of $P_r$ also corresponds just
to the highest $l$, that is, $l=15$ and $l=30$, respectively.

Let us finish this note with the conclusions.
The statistical complexity and the Fisher-Shannon information have been shown to
be independent of the potential strength, $\lambda$. It is the specific forms,
(\ref{eq:H}),(\ref{eq:D}) and (\ref{eq:J}),(\ref{eq:I}), of the definitions of these 
two indicators that yield this striking property.
This fact could be an indirect argument to justify the choice of these expressions.
Then, these magnitudes have been calculated.
We have taken advantage of the exact knowledge of the wave functions.
Concretely, we put in evidence that, for a fixed level of energy, let us say $e_{n,l}=2n+l$, 
these statistical magnitudes take their minimum values for the highest allowed 
orbital angular momentum, $l=e_{n,l}$. 
It is worth to remember at this point that the radial part of this particular wave function,
that describes the quantum system in the $(n=0,l=e_{n,l})$ orbital, has no nodes. This means
that the spatial configuration of this state is, in some way, a spherical-like shell. 
In the Appendix B, the mean radius of this shell, $<r>_{n,l,m}$, is found for the 
case $(n=0,l=e_{n,l},m)$. This is: 
\begin{equation}
<r>_{n=0,l=e_{n,l},m}\equiv <r>_{n=0,l=e_{n,l}} \simeq 
\,\sqrt{\lambda^{-1}\,(e_{n,l}+1)}\left(1+\Theta({e_{n,l}^{-1}})\right),
\label{eq:re}
\end{equation}
that tends, when $e_{n,l}\gg 1$, to the radius of the $Nth$ energy level, 
$r_{N}=\sqrt{\lambda^{-1}\,(N+1)}$, taking $N=e_{n,l}$ in the Bohr-like picture 
of the harmonic oscillator (see Appendix B). 

As it was remarked in Ref. \cite{sanudo2008}, here we also obtain that
the minimum values of the statistical measures calculated
from the wave functions of the quantum isotropic harmonic oscillator
select just those orbitals that in the pre-quantum image are the Bohr-like orbits. 
Therefore, we conclude that our intuition is enhanced when using these magnitudes 
to discern complexity at a quantum level.



\section*{APPENDIX A: Invariance of $C$ and $P$ under rescaling transformations} 

Here, we show that the statistical complexities $C_r$ and $C_p$ are equal 
and independent of the strength potential, $\lambda$,
for the case of the quantum isotropic harmonic oscillator.
Also, the same behavior is displayed by $P_r$ and $P_p$.

For a fixed set of quantum numbers, $(n,l,m)$,
let us define the normalized probability density $\hat{\rho}(\vec{t})$:
\begin{equation}
\hat{\rho}(\vec{t})= {2\;n!\over \Gamma(n+l+3/2)}\;
t^{2l}\;e^{-{t^2}}\; \left[L_{n}^{l+1/2}({t^2})\right]^2\;|Y_{l,m}(\Omega)|^2.
\end{equation}
From expressions (\ref{eq:Psi}), (\ref{eq:R}) and (\ref{eq:rho}),
it can be obtained that
\begin{equation}
\rho_{\lambda}(\vec{r})=\lambda^{3/2}\;\hat{\rho}(\lambda^{1/2}\vec{r}),
\end{equation}
where $\rho_{\lambda}$ is the normalized probability density of expression (\ref{eq:rho}). 
Now, it is straightforward to find that
\begin{equation}
H_r(\rho_{\lambda})= \lambda^{-3/2}\;H(\hat{\rho})\,,
\end{equation}
and that 
\begin{equation}
D_r(\rho_{\lambda})={\lambda^{3/2}\;D(\hat{\rho})}.
\end{equation}
Then,
\begin{equation}
C_r(\rho_{\lambda})=C(\hat{\rho}),
\end{equation} 
and the non $\lambda$-dependence of $C_r$ is shown.

To show that $C_r$ and $C_p$ are equal, let us note that,
from expressions (\ref{eq:Psih}), (\ref{eq:Rh}) and (\ref{eq:rho}),
the normalized probability density $\gamma_{\lambda}(\vec{p})$ for the same
set of quantum numbers $(n,l,m)$ can be written as
\begin{equation}
\gamma_{\lambda}(\vec{p})=\lambda^{-3/2}\;\hat{\rho}(\lambda^{-1/2}\vec{p}).
\end{equation} 
Now, it is found that
\begin{equation}
H_p(\gamma_{\lambda})=\lambda^{3/2}\;H(\hat{\rho})\,,
\end{equation}
and that 
\begin{equation}
D_p(\gamma_{\lambda})={\lambda^{-3/2}\;D(\hat{\rho})}.
\end{equation}
Then,
\begin{equation}
C_p(\gamma_{\lambda})=C(\hat{\rho}),
\end{equation} 
and the equality of $C_r$ and $C_p$, and its non $\lambda$-dependence
are shown.

Equivalently, from expressions (\ref{eq:P}), (\ref{eq:J}), (\ref{eq:Ir}) 
and (\ref{eq:Ip}), it can be found that $P_r=P_p$, and that these magnitudes
are also non $\lambda$-dependent.

\section*{APPENDIX B: Bohr-like orbits in the quantum isotropic harmonic oscillator} 

Here, the mean radius of the orbital with the lowest complexity is calculated 
as function of the energy. Also, the radii of the orbits in the Bohr picture are 
obtained.

The general expression of the mean radius of a state represented by the
wave function $\Psi_{n,l,m}$ is given by 
\begin{equation}
<r>_{n,l,m}\equiv <r>_{n,l}={n!\over \Gamma(n+l+3/2)}\;{1\over \lambda^{1/2}}\;
\int_0^{\infty} t^{l+1}\;e^{-{t}}\; \left[L_{n}^{l+1/2}({t})\right]^2\;dt.
\label{eq:r}
\end{equation}
For the case of minimum complexity (see Fig. 1 or 2),
the state has the quantum numbers $(n=0,l=e_{n,l})$. The last expression (\ref{eq:r}) becomes:
\begin{equation}
<r>_{n=0,l=e_{n,l}}={(e_{n,l}+1)!\over \Gamma(e_{n,l}+3/2)\lambda^{1/2}}\;,
\end{equation}
that, in the limit $e_{n,l}\gg 1$, simplifies to expression (\ref{eq:re}):
\begin{equation}
<r>_{n=0,l=e_{n,l}\gg 1}\simeq 
\,\sqrt{\lambda^{-1}\,(e_{n,l}+1)}\left(1+\Theta({e_{n,l}^{-1}})\right).
\label{eq:rn0}
\end{equation}

We proceed now to obtain the radius of an orbit in the Bohr-like image of the 
isotropic harmonic oscillator. Let us recall that this image establishes the quantization 
of the energy through the quantization of the classical orbital angular momentum.
So, the energy $E$ of a particle of mass $m$ moving with velocity $v$ on a circular 
orbit of radius $r$ under the harmonic potential $V(r)=m\lambda^2r^2/2$ is:
\begin{equation}
E={1\over 2}m\lambda^2r^2+{1\over 2}mv^2.
\label{eq:E1}
\end{equation}
The circular orbit is maintained by the central force through the equation:
\begin{equation}
{mv^2\over r}=m\lambda^2r.
\end{equation}
The angular momentum takes discrete values according to the condition:
\begin{equation}
mvr=(N+1)\hbar \hspace{1cm} (N=0,1,2,\ldots).
\label{eq:mv}
\end{equation}
Combining the last three equations (\ref{eq:E1})-(\ref{eq:mv}), 
and taking atomic units, $m=\hbar=1$, the radius $r_N$ of a Bohr-like orbit
for this system is obtained
\begin{equation}
r_N=\sqrt{\lambda^{-1}(N+1)} \hspace{1cm} (N=0,1,2,\ldots).
\end{equation}
Let us observe that this expression coincides with the quantum mechanical radius 
given by expression (\ref{eq:rn0}) when $e_{n,l}=N$ for $N\gg 1$.

\newpage

\begin{figure}[h]
\centerline{\includegraphics[width=7cm]{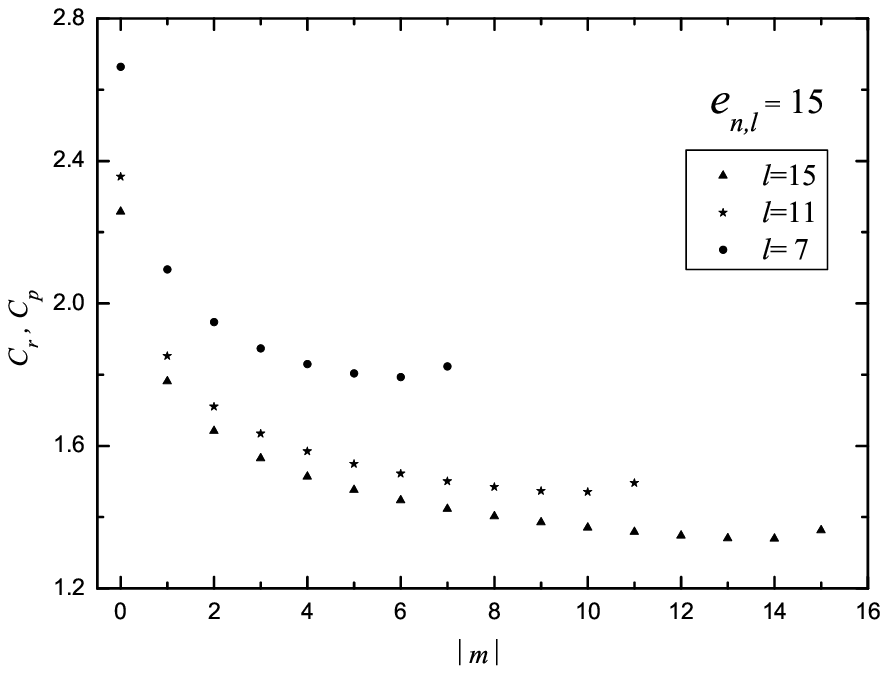}\hskip 5mm\includegraphics[width=7cm]{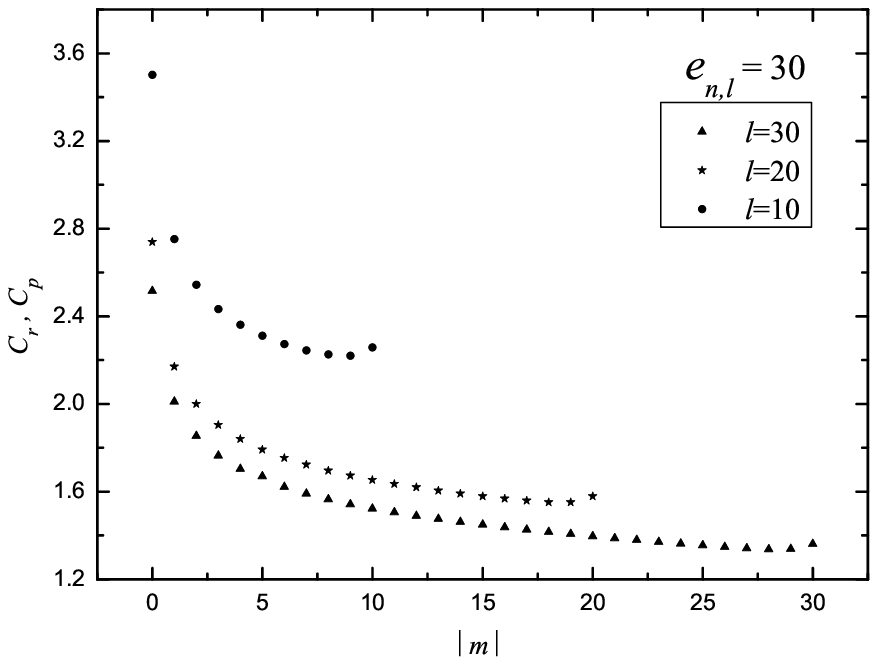}}
\centerline{(a)\hskip 7cm (b)} 
\caption{Statistical complexity in position space, $C_r$, and momentum space, $C_p$, vs. 
$|m|$ for different energy $e_{n,l}$-values in the quantum isotropic harmonic oscillator 
for (a) $e_{n,l}=15$ and (b) $e_{n,l}=30$. Recall that $C_r=C_p$. All values are in atomic units.}
\label{fig1}
\end{figure}

\vspace{2cm}
\begin{figure}[h]
\centerline{\includegraphics[width=7cm]{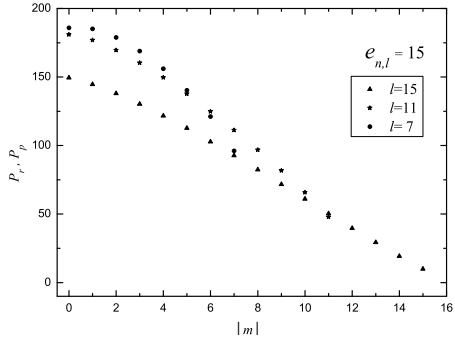}\hskip 5mm\includegraphics[width=7cm]{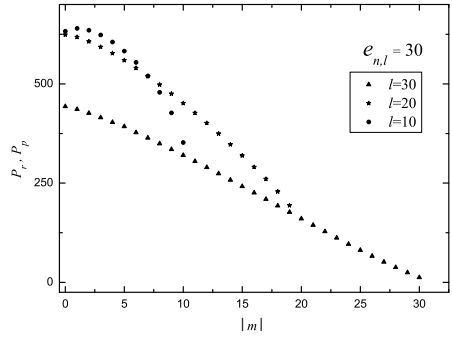}}
\centerline{(a)\hskip 7cm (b)} 
\caption{Fisher-Shannon information in position space, $P_r$, and momentum space, $P_p$, vs. 
$|m|$ for different energy $e_{n,l}$-values in the quantum isotropic harmonic oscillator. 
for (a) $e_{n,l}=15$ and (b) $e_{n,l}=30$. Recall that $P_r=P_p$. 
All values are in atomic units.} 
\label{fig2}
\end{figure}

\end{document}